\documentclass[english]{emulateapj}
\usepackage{amsmath,amssymb}
\usepackage[T1]{fontenc}
\usepackage[latin1]{inputenc}
\makeatletter

\include{epsf}

\begin{document}
\title{
Molecular Hydrogen Formation
on Ice Under Interstellar Conditions }
\author{Hagai B. Perets and
          Ofer Biham}
\affil{Racah Institute of Physics, The Hebrew University of Jerusalem,
        Jerusalem 91904, Israel}
\author{Giulio Manic\'o  and Valerio Pirronello}
\affil{Dipartimento di Metodologie Fisiche e Chimiche per l'Ingegneria, \\
        Universita' di Catania, 95125 Catania, Sicily, Italy}
\and
\author{Joe Roser, Sol Swords
        and Gianfranco Vidali}
\affil{Department of Physics, Syracuse University,
        Syracuse, NY 13244}

\begin{abstract}

The results of experiments on the formation of molecular hydrogen on
low density and high density
amorphous ice surfaces are analyzed using a rate equation model. 
The activation energy barriers for the relevant diffusion and
desorption processes are obtained.
The more porous morphology of the low density ice gives rise
to a broader spectrum of energy barriers compared to the
high density ice.
Inserting these parameters into the rate equation model under
steady state conditions, we evaluate the production rate of 
molecular hydrogen on ice-coated interstellar dust grains. 

\end{abstract}

\keywords{dust--- ISM; abundances --- ISM; molecules --- molecular processes}

\newpage
 
\section{Introduction}
\label{intro.}

The formation of molecular hydrogen in the interstellar medium (ISM)
is a process of fundamental importance
in astrophysics because H$_2$ helps the initial cooling of clouds
during gravitational collapse and enters, either in neutral or ionized form, 
most reaction schemes that make other molecules 
\citep{Duley1984,Williams1998}.  
It was recognized long ago
that H$_2$ cannot form in the gas phase
efficiently enough to account for its observed abundance
\citep{Gould1963}.  
It was proposed that
molecular hydrogen formation takes place on dust grains that act as
catalysts allowing the protomolecule to quickly release the 4.5 eV of
excess energy in a time comparable to the vibration period of the
highly vibrationally excited state in which it is formed.

The process of H$_2$ formation on grains
can be broken up into a few steps as follows.
An H atom approaching the surface of a grain has
a probability $\xi$ to become adsorbed.
The adsorbed H atom (adatom)
resides on the surface for an average time 
$t_{\rm H}$ (residence time)
before it desorbs.
In the Langmuir-Hinshelwood mechanism,
the adsorbed H atoms quickly equilibrate with the surface
and diffuse on the surface of the grain
either by thermal activation or tunneling. 
Furthermore, atoms that are deposited on top of already adsorbed atoms
are rejected.
When two adsorbed H atoms encounter each other, 
an H$_2$ molecule may form with a certain probability
\citep{Williams1968,Hollenbach1970,Hollenbach1971a,Hollenbach1971b,Smoluchowski1981,Aronowitz1985,Duley1986,Pirronello1988,Sandford1993,Takahashi1999,Farebrother2000}.

The sticking probability of
H atoms on dust grains was calculated using
semi-classical methods.
Quantum mechanical calculations of the mobility of H
atoms on a polycrystalline water ice surface,
showed that tunneling between adsorption sites,
would provide H adatoms with sufficient mobility to guarantee efficient
H$_2$ formation
even at 10 K
\citep{Hollenbach1970,Hollenbach1971a,Hollenbach1971b}.
The steady state production rate of molecular hydrogen,
$R_{\rm H_2}$ (cm$^{-3}$ s$^{-1}$)
was expressed by

\begin{equation}
\label{eq:salpeter}
    R_{{\rm H_2}} = {1 \over 2}
        n_{{\rm H}} v_{{\rm H}} \sigma \gamma n_{\rm g},
\end{equation}

\noindent
where
$n_{{\rm H}}$ (cm$^{-3}$)
and
$v_{{\rm H}}$ (cm s$^{-1}$)
are the number density and the speed
of H atoms in the gas phase, respectively,
$\sigma$ (cm$^{2}$) 
is the average cross-sectional area of a grain and
$n_{{\rm g}}$ (cm$^{-3}$)
is the number density of dust grains.
The parameter
$\gamma$ is the fraction of H atoms striking the grain
that eventually form a molecule, namely $\gamma = \xi \eta$,
where
$\eta$
is the probability that an H adatom on the surface
will recombine with another H atom
to form
H$_2$.
The probability $\xi$ for an H atom to become adsorbed on a grain surface
covered by an ice mantle has been calculated by 
Buch and Zhang (1991) 
and 
Masuda et al. (1998).
They found that $\xi$ depends on the surface
temperature and on the energy of the irradiation beam.
For a surface at 10 K and beam temperature of 350 K,
Masuda et al. (1998) obtained a sticking coefficient around 0.5.

Alternative mechanisms, which were shown to operate in selected
experiments on well characterized crystalline surfaces, are the
Eley-Rideal mechanism and the hot-atom mechanism. 
In the Eley-Rideal mechanism, an
atom from the gas phase impinges on 
another atom, which is already adsorbed,
and reacts
with it before becoming thermally accommodated with the surface. 
This mechanism has been verified experimentally for
D atoms on a Cu surface 
\citep{Rettner1996}.
In the hot-atom mechanism 
\citep{Harris1981}, 
an atom lands on the surface and moves
from site to site at super-thermal speed 
until it reacts with another atom 
(before it thermally accommodates with the surface). 
In this case, the atom might fail to give up
all of its incoming kinetic energy, or it might use some of the energy
gained in becoming confined to the surface to move across it at
super-thermal energy. The two mechanisms can be distinguished
experimentally by measuring the cross-section for the reaction or the
energy carried away from the just-formed molecule 
\citep{Zecho2002}.
The kinetic energy, as well as the vibrational and rotational
states of the desorbed molecule, are expected to depend on
the surface mechanism through which it was formed.

In the last few years, we have conducted a series of experiments to
study the formation of molecular hydrogen on dust grain analogues under
conditions relevant to astrophysical environments
\citep{Pirronello1997a,Pirronello1997b,Pirronello1999,Roser2002,Roser2003}.
The experiments used well collimated beams of hydrogen and deuterium
atoms. The production of HD that occurs on the surface of a dust grain
analogue was measured both during the irradiation with the beams and
during a subsequent temperature programmed desorption 
(TPD) experiment. 
In this case, the temperature of the sample is raised
quickly to either desorb particles (atoms and molecules) 
that got trapped on the surface or
to enhance their diffusion and 
subsequent reaction and/or desorption. 
In order to disentangle the process of diffusion from the one of
desorption, separate experiments were carried out in which molecular species
were irradiated on the sample and then were induced to desorb.

In an earlier set of experiments, the formation of molecular hydrogen
was studied on samples of polycrystalline olivine and amorphous carbon
\citep{Pirronello1997a,Pirronello1997b,Pirronello1999}
and the results were analyzed
using rate equation models 
\citep{Katz1999,Cazaux2002,Cazaux2004}.
In this analysis
the parameters of the rate equations
were fitted to the experimental TPD curves.
These parameters are
the activation energy barriers for atomic
hydrogen diffusion and desorption, 
the barrier for molecular hydrogen
desorption 
and the fraction of
molecules that desorb upon recombination.  
Using the values of the
parameters that fit best the experimental results, 
the efficiency of
hydrogen recombination on the polycrystalline 
olivine and amorphous carbon surfaces 
was calculated
for interstellar conditions.
By varying the temperature and flux over the astrophysically relevant range, 
the domain in which 
there is non-negligible recombination efficiency
was identified. 
It  was found that the
recombination efficiency is highly temperature dependent. 
For each of the two samples 
there is a narrow window of high efficiency along the temperature axis,
which shifts to higher temperatures as the flux is increased. 
For the astrophysically relevant flux range the
high efficiency temperature range for polycrystalline olivine 
was found to be between
$7 - 9$K,  
while for amorphous carbon it is between
$12 - 16$ K.

More recently, these studies were extended to different forms of water
ice, namely high density ice (HDI) and low density ice (LDI) 
\citep{Manico2001,Roser2002,Hornekaer2003}. 
In these experiments
it was seen that the type of amorphous ice affects both 
the kinetics of molecular hydrogen formation and its efficiency.
We then carried out a
detailed analysis of the new data using a rate equation model in order
to obtain useful physical parameters for the modeling of
chemistry in dense interstellar clouds.
Specifically, the parameters of the rate equation
model are fitted to the experimental TPD curves.
These parameters are
the activation energy barriers for diffusion and desorption
of hydrogen atoms and hydrogen molecules
as well as the population ratios of different types of
adsorption sites for molecular hydrogen.
Using the values of the
parameters that fit best the experimental results,
the efficiency of
hydrogen recombination on the amorphous ice surfaces
is obtained for
values of the hydrogen flux
and the surface temperature
pertinent
to a range of interstellar conditions.

The paper is organized as follows.
In Sections
\ref{sec:Experimental} and
\ref{sec:ExpResults}
we describe the experiments to be analyzed and their results.
The rate equation models are
introduced in Sec.
\ref{sec:Model}.
Subsequent analysis and results are
presented in Sec.
\ref{sec:analysis},
followed by a discussion
in Sec.
\ref{sec:discussion}
and a summary in Sec.
\ref{sec:Summary}.

\section{Review of Experimental Methods}
    \label{sec:Experimental}
The experimental apparatus and measurement techniques
were most recently described in
Vidali et al. (2004).
Here we give a brief outline.
The apparatus  consists of
an ultra-high vacuum (UHV) chamber pumped by a cryopump and a
turbo-molecular pump
(operating pressure in the low 10$^{-10}$ Torr range).
The sample is placed in the center of the UHV chamber and is
mounted on a
rotatable UHV compatible liquid
helium continuous flow cryostat.
By varying the flow of liquid helium and by using a heater located 
behind the sample, it is possible to change the temperature of
the sample
over a wide range.
In the thermal desorption experiments, the temperature is
raised quickly from around 10 K to about 30 K. Higher temperatures are
obtained at slower rates, as when the sample temperature is raised to
$\sim$ 90 K
to change the structure of the ice (see below) or to desorb the
ice layers. 
The temperature
is measured by an iron-gold/chromel thermocouple and a calibrated
silicon diode placed in contact with the sample.

Atoms are sent to the surface of the sample via
two triple
differentially pumped atomic beam lines aimed at this surface. 
Each has a radio-frequency cavity in which the molecules
are dissociated, optionally 
cooled to $\sim$ 200 K by passing the atoms
through a cold channel, and then injected into the
line. Dissociation rates are typically in the 75 to 90$\%$ range, and
are constant throughout a run.
Estimated fluxes are as low as 10$^{12}$ 
(atoms cm$^{-2}$s$^{-1}$)
\citep{Roser2002}.

The experiment is done in two phases. 
First, beams of H and D atoms are sent onto the surface,
at a constant irradiation rate $F_0$,
for a
given period of time 
$t_0$
(from tens of seconds to tens of minutes)
while the surface
temperature is maintained at a constant value $T_0$.
At this time any HD formed and released is detected
by a quadrupole mass spectrometer mounted on a rotatable 
flange in the main UHV chamber.
In the second phase,
starting at $t_0$ 
(the TPD phase), 
the irradiation is stopped and the
sample temperature is quickly 
ramped 
at a rate $b$
($\sim$ 0.6 K/sec on average), 
and the HD
signal is measured. 
The time dependence of the surface temperature, $T(t)$, during
typical TPD experiments is shown in Fig. 1,
and is approximated by a piecewise-linear fit.
In the ideal case in which the heating rate is nearly constant,
the time dependence of the flux and surface temperature 
can be approximated by

\begin{mathletters}
\begin{eqnarray}
\label{eq:temp}
F(t) & = & F_{0}; \ \ \ \ \ \  T(t)  =  T_{0}:
\ \ \ \ \ \ \ \ \ \ \ \ \ \ \ \ \ \ \ 0 \leq t < t_{0}
\label{eq:temp1} \\
F(t) & = & 0;    \ \ \ \ \ \ \ T(t)  =  T_{0} + b (t - t_{0}):
\ \ \ \ \  t \geq t_{0}.
\label{eq:temp2}
\end{eqnarray}
\end{mathletters}

The HDI sample is prepared by directly depositing a measured quantity of
water vapor via a stainless steel capillary placed in the proximity
of the sample (a polished copper disk). 
Before deposition, the water undergoes
repeated cycles of freezing and thawing to remove trapped gases. The
sample is held at 10 K or lower during deposition and the deposition
rate is ~8 layers/sec, for a total thickness of the order of 1200
layers. This method of preparing the ice sample should produce a high
density ice, or HDI 
\citep{Jenniskens1994}.  
The other phase, namely the
low density amorphous ice or LDI, 
can be obtained from HDI by heating it.
This transformation is gradual
but irreversible
and occurs over a broad temperature range
starting from about 38 K. 
In our experiments, LDI
is prepared by gradually heating the HDI to 90 K and
holding the temperature at that value for at least 5 minutes. Then the
sample is cooled down to 10 K.
In the experiments analyzed here, the LDI sample was kept at a temperature
between 9 and 11 K during the irradiation phase. During the TPD phase, care
was exercised so that the sample didn't reach a temperature close to or
higher than 38 K
\citep{Roser2002,Vidali2004}.
Low and high density amorphous ice were shown to have similar
characteristics and structure as interstellar ice mantles
\citep{Mayer1986,Jenniskens1995}. 
The amorphous ice structure, which
contains many pores with diameters ranging between ~15-20 \AA, can
adsorb H$_{2}$ molecules, that can diffuse into ice micro-pores
where they can be trapped
\citep{Mayer1986,Langel1994}. 

\section{Experimental Results}
        \label{sec:ExpResults}

Irradiations with beams of H and D ("H+D" thereafter)
were done both on
LDI and HDI in order to explore
the formation processes of HD molecules on these amorphous ice surfaces. 
In a separate set of experiments, beams of HD and D$_2$ molecules
were irradiated on the ice surfaces. The latter experiments do
not involve formation of molecules.
However, they allow us to isolate and better
analyze some of the
parameters relevant to the molecular formation processes
explored directly in the former experiments. 
They also
support our hypothesis that molecules that consist of different hydrogen
isotopes behave quite similarly (although not exactly the same) 
as HD molecules and all of them could be 
treated by the same general rate equation model.

Using the methods described in the previous Section,
the following experiments were carried out.
The H+D irradiation runs were performed
with different irradiation times 
(2, 4, 8, 12 and 18 minutes on LDI; 
6, 8 and 18 minutes on HDI).
The
surface temperatures 
during irradiation were
$T_{0} \simeq 9.5$ K 
in the LDI experiments 
and
$T_{0} \simeq 14.5$ K
in the HDI experiments. 
In the experiments with molecular beams
the irradiation 
time was of 4 minutes.
The surface temperatures
during irradiation of
HD and D$_2$
on LDI
were
$T_{0} \simeq 9$ K 
and 
$T_{0} \simeq 10$ K,
respectively.
In Table 1 we present a list of the experimental runs analyzed
in this paper.
During the TPD runs, the sample temperature 
is monitored as a function of time. 
The time dependence of the sample temperature during typical
temperature ramps 
is shown in Fig. 1 for
LDI (solid line)
and
HDI (dashed line).
The temperature ramps deviate from linearity but they are highly
reproducible.
The symbols show the experimental measurements
and the lines are piecewise-linear fits.
For the LDI experiments a good fit is obtained with
two linear segments, with an average heating rate of 
$b \simeq 0.6$ K/sec over the whole ramp.
For the HDI experiments a good fit is obtained
with three linear segments, with an average heating
rate of 
$b \simeq 0.13$ K/sec.
Although the temperature ramps deviate from linearity, most of
the structure of the TPD peaks takes place within a temperature
range in which it is linear to a very good approximation. 
The analysis of the data is done using these temperature ramps,
thus any possible effect of the deviation from linearity is
taken into account. Such deviations are found to be minor.

The desorption rates of molecules vs. surface temperature during
the TPD runs are 
shown in Figs. 2-7.
In Fig. 2 we
present the desorption rate of HD molecules after
irradiation by
HD and H+D on LDI.
In both cases the TPD curves are broad. In the case
of HD irradiation, three peaks can be identified
in the TPD curve.
For H+D irradiation there are only two peaks.
Interestingly, they coincide on the temperature
axis with two of the three peaks obtained for 
HD irradiation.
In Fig. 3 we show the
desorption rate of HD molecules after 
irradiation by H+D on LDI for several irradiation 
times between 2 minutes and 18 minutes.
As in Fig. 2, all these TPD curves exhibit two peaks.
The position of the high temperature peak is found to
be independent of the irradiation time, indicating that
this peak exhibits first order kinetics.
The low temperature peak shifts to the right as the
irradiation time decreases, thus, this peak
exhibits second order kinetics.
Fig. 4 shows the desorption rate of D$_2$ molecules after
D+D and D$_2$
irradiation on LDI.
These TPD curves are also broad, and are qualitatively
similar to those of HD desorption, shown in Fig. 2. 

In Fig. 5 we show 
the desorption rate of HD molecules after irradiation by
HD and H+D on HDI.
Here the TPD curves are narrower and include only one
peak. 
In Fig. 6 we show the desorption rate of HD molecules after
irradiation by H+D on HDI for three irradiation times.
The peak shifts to the right as the
irradiation time decreases, namely, it 
exhibits second order desorption.
Fig. 7 shows
the desorption rate of D$_2$ molecules after
D$_2$ irradiation on HDI.
A single peak is observed, qualitatively similar to the
one shown in Fig. 5 for HD.

In the experiments in which beams of H+D atoms were
irradiated on LDI and HDI samples, 
most of the HD molecules detected were formed 
during the heat pulse.
Only a small fraction of them
were formed during the irradiation process
\citep{Roser2002}.
This indicates
that at least under our experimental conditions, 
prompt-reaction mechanisms
\citep{Duley1986}
or fast tunneling
\citep{Hollenbach1971b}
do not play a major role in the formation of molecules.

\newpage
\section{The Rate Equation Models}
    \label{sec:Model}

In the TPD curves studied here most of the adsorbed hydrogen is
released well before a temperature of 40 K is reached.
We thus conclude that
the hydrogen atoms on the surface are trapped in
physisorption potential wells and are only weakly adsorbed.
The mechanism for the formation of
H$_2$ 
(as well as HD and D$_2$) is assumed to be
the Langmuir-Hinshelwood (LH) scheme. 
In this scheme, 
the rate of formation
of molecular hydrogen is diffusion limited.
This assumption is justified
by the Langmuir-like behavior 
observed in the desorption curves.

The analysis of thermal programmed desorption
experiments usually starts with the Polanyi-Wigner expression for the
desorption rate R(t):

\begin{equation}
R(t) = \nu N(t)^{\beta} \exp(- E_d / k_B T),
\label{eq:desorption}
\end{equation}
\noindent
where $N$ is the coverage of
reactants on the surface,
$\beta$ is the order of desorption,
$\nu$ is the attempt frequency,
$E_d$ is the effective
activation energy for the dominant recombination and desorption process
and $T=T(t)$ is the sample temperature.
In the TPD experiment,
first order
($\beta=1$)
desorption curves
$R(t)$
exhibit asymmetric peaks with a sharp
drop-off on the high temperature side.
The position of the peak is insensitive to 
the initial
coverage, determined by the irradiation time.
Second order desorption curves ($\beta=2$)
exhibit symmetric peak shapes.
These peaks shift toward
{\em lower temperatures}
as the initial coverage is increased
\citep{Chan1978}.
An important assumption is that
all activation energy barriers are coverage independent.
This assumption may not apply at high coverage.
However, at the low coverages obtained in the experiments
analyzed here (up to $\sim 1\% $ of a layer),
it is a reasonable assumption.

We introduce two models for describing the TPD curves: a simplified model
which succeeds in describing the main characteristics of the TPD curves and
a more complete model which describes them using a more general and accurate
view.

\subsection{The Simple Model}
\label{sec:Simple Model}

Consider an experiment in which a flux of H atoms
is irradiated on the surface.
H atoms that stick to the
surface hop as random walkers.
The hopping atoms may either encounter each other and form H$_2$
molecules, or desorb from the surface. 
As the sample temperature is raised, both the hopping
and desorption rates quickly increase.
In the models used here, there is no distinction between the H and 
D atoms, namely 
the same diffusion and desorption barriers are used for
both isotopes.

In the models we assume a given density of adsorption sites
on the surface. Each site can adsorb either an H atom
or an H$_2$ molecule. 
In terms of the adsorption of H atoms, 
all the adsorption sites are assumed to be identical,
where the energy barrier for H diffusion is
$E_{\rm H}^{\rm diff}$
and the barrier for desorption
is
$E_{\rm H}^{\rm des}$.
As for the adsorption of H$_2$ molecules, we assume that
the adsorption sites may differ from each other.
In particular, we assume that the population of adsorption
sites is divided into
$J$ types, where a fraction $\mu_j$ of the sites belong
to type $j$, where $j=1,\dots,J$,
and 
$\sum_j \mu_j = 1$.
The energy barrier for desorption of H$_2$ molecules
from an adsorption site of type $j$ is
$E_{\rm H_2}^{\rm des}(j)$.

This model is motivated by the
TPD curves obtained after irradiation by HD and D$_2$
molecules on LDI, shown in Figs. 2 and 4, respectively.
These TPD curves
are broad and can be divided into three peaks. 
We interpret this feature as an indication that there are
three types of adsorption sites for molecules, 
which differ from each other in the 
energy barriers for desorption.

Let $N_{\rm H}$ 
[in monolayers ($ML$)] be the coverage
of H atoms on the surface, namely the
fraction of adsorption sites that are
occupied by H atoms. 
Similarly, let
$N_{\rm H_2}(j)$ (also in $ML$) be
the coverage of H$_{2}$ molecules that are 
trapped in adsorption sites of type $j$, where 
$j=1,\dots,J$.
Clearly, this coverage is limited by the number of
sites of type $j$ and therefore
$N_{\rm H_2}(j) \le \mu_j$.
Since we assume that each site can host only one atom or one molecule,
the coverage cannot exceed a monolayer, and thus
$N_{\rm H} + \sum_j N_{\rm H_2}(j) \le 1$. 

Our analysis shows that for LDI one needs three types of
molecular adsorption sites, namely $J=3$, while for 
HDI good fits are obtained with $J=1$.
For the case of LDI we thus 
obtain the following set of rate equations
\begin{mathletters}
\begin{eqnarray}
& \dot{N}_{{\rm H}} & =  
F\left[1 - N_{\rm H} - \sum_{j=1}^{3}{N_{\ H_2}(j)}\right] \nonumber \\ &-&
W_{\rm H}N_{\rm H} - 2\alpha N_{\rm H}^{2} 
\label{eq:NH} \\
& \dot{N}_{\rm H_2}(1) & 
= \mu_{1} \alpha {N_{\rm H}}^{2} - W_{\rm H_2}(1)N_{\rm H_2}(1)
\label{eq:N2(1)} \\
& \dot{N}_{\rm H_2}(2) & = 
\mu_{2} \alpha {N_{\rm H}}^{2} - W_{\rm H_2}(2)N_{\rm H_2}(2)
\label{eq:N2(2)}\\
& \dot{N}_{\rm H_2}(3) & = 
\mu_{3} \alpha {N_{\rm H}}^{2} - W_{\rm H_2}(3)N_{\rm H_2}(3)
\label{eq:N2(3)}
\end{eqnarray}
\end{mathletters}

\noindent
The first term on the right hand side of
Eq.~(\ref{eq:NH})
represents the incoming
flux in the Langmuir kinetics.
In this scheme H atoms deposited on top of H atoms
or H$_2$ molecules already on the surface are rejected.
$F$ represents an
{\em effective} flux (in units of $ML$/sec),
namely it
already includes the possibility of a temperature
dependent sticking coefficient.
In practice we find that for the conditions studied
here the Langmuir rejection term is negligible and it
is thus ignored in the simulations.
The second term in
Eq.~(\ref{eq:NH})
represents the desorption of H atoms from the
surface.
The desorption coefficient is

\begin{equation}
W_{H} =  \nu \exp (- E_{\rm H}^{\rm des} / k_{B} T)
\label{eq:Pdes}
\end{equation}

\noindent
where $\nu$ is the attempt rate
(standardly taken to be $10^{12}$ $s^{-1}$),
$E_{\rm H}^{\rm des}$
is the activation energy barrier for desorption
of an H atom and $T$ is the temperature.
The third term in
Eq.~(\ref{eq:NH})
accounts for the depletion of the H population
on the surface due to diffusion-mediated recombination 
into H$_{2}$ molecules,
where

\begin{equation}
\alpha =  \nu \exp (- E_{\rm H}^{\rm diff} / k_{B} T)
\label{eq:Alpha}
\end{equation}

\noindent
is the hopping rate of H atoms on the surface
and $E_{\rm H}^{\rm diff}$ 
is the activation energy barrier for H diffusion.
Here we assume that there is no barrier for recombination. 
If such a barrier
is considered, it can be introduced as discussed in 
Pirronello et al. (1997b,
1999).

Eqs.~(\ref{eq:N2(1)})-(\ref{eq:N2(3)}) 
describe the population of molecules on the surface.
The first term on the right hand side of
each of these three equations
represents the formation of H$_2$ molecules
that become adsorbed in a site of type 
$j=1$, 2 or 3.
The second term in
Eqs.~(\ref{eq:N2(1)})-(\ref{eq:N2(3)})
describes the desorption of H$_2$ molecules from
sites of type $j$,
where

\begin{equation}
W_{\rm H_2}(j)  =  
\nu \exp (- E_{\rm H_2}^{\rm{des}}(j) / k_{B} T),
\label{eq:Wk}
\end{equation}

\noindent
is the H$_2$ desorption coefficient and
$E_{\rm H_2}^{\rm{des}}(j)$ 
is the activation energy
barrier for H$_2$ desorption from an adsorption site
of type $j$.
The H$_2$ production rate $R$ is given by:

\begin{equation}
R  =  \sum_{j=1}^{3}{W_{\rm H_2}(j) N_{\rm H_2}(j)}.
\label{eq:Production}
\end{equation}
\noindent

\noindent
This model can be considered as a generalization of
the Polanyi-Wigner model
[see
Eq.
(\ref{eq:desorption})].
It gives rise to a wider
range of applicability,
compared to
Eq.
(\ref{eq:desorption}).
In particular, it
describes both first order
and second order
desorption kinetics
(or a combination)
for different regimes of temperature and flux
\citep{Biham1998}.

Experiments that involve irradiation by molecules are useful
since they enable an independent evaluation of the parameters
of molecular adsorption on the surface.
Consider the case where a flux of H$_2$ molecules is 
irradiated on the surface.
In this case 
Eq.~(\ref{eq:NH})
is no longer relevant. 
The H$_2$ molecules on the surface
are distributed
between the different types of adsorption sites,
in proportion to the
population ratios $\mu_j$.
Each one of Eqs.~(\ref{eq:N2(1)})-(\ref{eq:N2(3)})
can now be treated as independent
and solved analytically.
The TPD curve consists of
three first order peaks, at temperatures determined by 
the desorption barriers 
$E_{\rm H_2}^{\rm des}(j)$, $j=1,2,3$. 
The sizes of these peaks are determined by
the population ratios $\mu_{j}$.
Given a TPD curve, 
the energy barriers 
$E_{\rm H_2}^{\rm des}(j)$ 
can be obtained from the 
temperatures 
$T_{\rm max}(j)$
of the corresponding peaks.
Solving for the maximal desorption rate
by taking its derivative with respect to the
temperature we obtain

\begin{equation}
\label{eq:Tmax}
{bE_{\rm H_2}^{\rm{des}}(j)\over{k_{B}{T_{\rm max}(j)^{2}}}} 
= {\nu \exp\left[{-E_{\rm H_2}^{\rm{\rm des}}(j)
   \over{k_{B}T_{\rm max}(j)}}\right]}.
\end{equation}

\noindent
In general, the parameters $\mu_{j}$, $j=1,\dots,J$ 
can be obtained by fitting the TPD curve as a sum of 
$J$ Gaussians.
Each of the $\mu_j$'s is obtained as the ratio between
the area below the corresponding Gaussian and the 
total area below the TPD curve.
In the case of HDI such procedure is not needed because
the TPD curve can be fitted with a single type of 
molecular adsorption site.

\subsection{The Complete Model}
\label{sec:Complete Model}

The simple model presented above 
provides good fits to the results of
the experiments in which H+D, HD and D$_2$
were irradiated on HDI (Figs. 5-7).
In this case it was assumed that there is only a
single type of molecular adsorption site.
In experiments on LDI, 
where several types of molecular adsorption sites are assumed,
the fitting of the TPD curves 
can be improved by using a more complete model. 
In this model the
H$_2$ molecules are allowed to diffuse between the 
different types of adsorption sites.
For the case of LDI (J=3),
the complete model takes the form

\begin{mathletters}
\begin{eqnarray}
\label{eq:CNH}
&\dot{N_{\rm H}} & =  
F\left[1 - N_{\rm H} - \sum_{j=1}^{3}{N_{\rm H_2}(j)}\right] \nonumber\\&-&
W_{\rm H}N_{\rm H} - 2\alpha N_{\rm H}^{2}
\\
&\dot{N}_{\rm H_2}(1) & =  {\mu_{1}} \alpha {N_{\rm H}}^{2} 
- \alpha_{\rm H_2}(1)N_{\rm H_2}(1) 
\nonumber\\&-& W_{\rm H_2}(1)N_{\rm H_2}(1) 
+ 
{\mu_{1}}\sum_{j=1}^{3}{{\alpha_{\rm H_2}(j)N_{\rm H_2}(j)}})
\label{eq:CN2(1)} \\
&\dot{N}_{\rm H_2}(2) & =  {\mu_{2}} \alpha {N_{\rm H}}^{2} 
- \alpha_{\rm H_2}(2)N_{\rm H_2}(2) \nonumber
\\&-& W_{\rm H_2}(2)N_{\rm H_2}(2) 
+
{\mu_{2}}\sum_{j=1}^{3}{{\alpha_{\rm H_2}(j)N_{\rm H_2}(j)}})
\label{eq:CN2(2)}\\
&\dot{N}_{\rm H_2}(3) & =  \mu_{3} \alpha {N_{\rm H}}^{2} 
- \alpha_{\rm H_2}(3)N_{\rm H_2}(3) \nonumber
\\&-& W_{\rm H_2}(3)N_{\rm H_2}(3) 
+
\mu_{3}\sum_{j=1}^{3}{\alpha_{\rm H_2}(j)N_{\rm H_2}(j)}).
\label{eq:CN2(3)}
\end{eqnarray}
\end{mathletters}

\noindent
The first equation is identical to the corresponding one
in the simple model, namely Eq.~(\ref{eq:NH}).
The first and third terms on the right hand side of
Eqs.~(\ref{eq:CN2(1)})-(\ref{eq:CN2(3)}) are the same as
those that appear in 
Eqs.~(\ref{eq:N2(1)})-(\ref{eq:N2(3)})
of the simple model. 
The second and fourth terms describe the diffusion 
of H$_2$ molecules between
adsorption sites of different types,
where
$\alpha_{\rm H_2}(j) = \nu \exp{(-E_{\rm H_2}^{\rm diff}/k_B T)}$
and $E_{\rm H_2}^{\rm diff}$
is the activation energy for hopping of an
H$_2$ molecule out of a site of type $j$.
The second terms account for the hopping of 
molecules out of adsorption sites of type $j$,
while the fourth terms account for their 
re-distribution between adsorption sites of the three
different types.

\section{Analysis of Experimental Results}
\label{sec:analysis}

In this Section we use the rate equation models to
fit the experimental results, and thus obtain the
parameters that describe the diffusion and desorption
of hydrogen atoms and molecules on ice surfaces.
We use a combination of analytical and numerical tools.
In particular, the numerical integration of
the rate equation models is done using the 
Runge-Kutta stepper.
The result of the integration is a set of
TPD curves that are a function of the
chosen set of parameters, 
some of which can be determined analytically
from the experimental TPD curves.
The data obtained from each
TPD run includes
the time dependence of the flux
$F(t)$
and temperature
$T(t)$.
The temperature $T(t)$ is measured directly via a thermocouple.
The flux $F(t)$
($ML$/sec)
is estimated as described elsewhere
\citep{Vidali1998}.
An approximate value for
$F(t)$,
in the required units of $ML$/sec,
can be obtained by integrating the TPD spectra,
generating the total {\it yield}
of the various experiments. 
The flux is then obtained from
the exponential fit indicated by Langmuir kinetics.
It is important to stress that
this is a lower bound value for the flux, and this value is reached
only if there is no desorption 
of H atoms during the TPD runs.
In the analysis we assume
that each of the energy barriers 
$E_{\rm H_2}^{\rm des}(j)$, $j=1,2,3$, 
represents a
Gaussian distribution
of energy barriers centered around it. 
Although the Gaussian distribution 
provides much better fits,
the important features of the curves depend
only on the parameters mentioned above. 
The Gaussian distribution is
inserted into the equations by 
effectively 
splitting each of the Eqs.
(\ref{eq:CN2(1)})-(\ref{eq:CN2(3)})
into about ten equations, 
each with a different 
energy barrier for H$_2$ desorption
according to the Gaussian distribution
around its central value. 

\subsection{Analysis of Experiments with Irradiation of Molecules}
\label{subsec:Molecules Irradiation}

The TPD curves obtained after 
irradiation of HD and D$_2$ on HDI
are narrow and exhibit a single peak
(Figs.~5 and 7).
In contrast, the TPD curves obtained after irradiation
of those molecules on LDI are broad and three distinct
peaks can be identified in them
(Figs. 2 and 4).
Since no surface reactions take place in these experiments,
these peaks are of first order kinetics.
Thus, one can use 
Eq.~(\ref{eq:Tmax}) 
to obtain the activation energy
$E_{\rm H_2}^{\rm des}(j)$ for the desorption
of H$_2$ atoms that are trapped in an adsorption site
of type $j$ from the surface.
This calculation is done in the context of the simple model.

Using the complete model
in the case of LDI, with $j=1,2,3$,
does not change the positions of the three peaks. 
The complete model allows diffusion of molecules between 
different types of adsorption sites. 
As a result, there is a flow of molecules from the relatively shallow
adsorption sites towards the deeper ones.
Thus, the relative sizes of the three sub-populations of molecules
adsorbed on LDI become temperature dependent.

The fitting curves for the HDI experiments
with HD and D$_2$ are shown in 
Figs. 5-7.
Some uncertainty exists
in the value of the energy barrier for HD 
desorption from HDI.
The value 
$E_{\rm H_2}^{\rm des}(1)=68.7$ meV
was obtained from the experiments in which the irradiation
times were relatively short (Fig. 5).
The experiments in which the irradiation time was long
give rise to the somewhat lower value of
$E_{\rm H_2}^{\rm des}(1)=65.5$ meV (Fig 6).
Additional uncertainty arises due to the choice of the
choice of the attempt frequency $\nu$.
A deviation by an order of magnitude in its value
would modify the resulting energy barriers by 3-4 meV. 

In the case of LDI, where three types of adsorption sites can
be identified, the situation is more complicated.
The energy barriers for desorption of H$_2$ molecules from these
three types of adsorption sites are obtained from
Eq.~(\ref{eq:Tmax}). 
The population ratios
$\mu_j$, $j=1,2,3$ are
obtained from fitting the TPD curves using the rate
equations.
These TPD curves are fitted very well using the complete
rate equation model
(Figs. 2 and 4).
Within this paremetrization, a simplifying assumption used here is
that the hopping rate 
$\alpha_{\rm H_2}(j)$
out of an adsorption site of type $j$
is linearly proportional to the desorption rate
$W_{\rm H_2}(j)$
from the same site.
The experimental data is best fitted when the ratio between
these two barriers is around 1,
namely 
$\alpha_{\rm H_2}(j) = W_{\rm H_2}(j)$.
Assuming the same attempt frequency for the two 
processes this indicates that
$E_{\rm H_2}^{\rm diff} = E_{\rm H_2}^{\rm des}$.
This is an unusual result, since typically the barriers for
desorption are higher than the barriers for diffusion.
However, it may indicate that in the porous LDI the dominant
diffusion process is, in fact, desorption and re-adsorption
on the surface within the pores.
Note that the energy barriers for desorption of molecules are
obtained directly using Eq.
(\ref{eq:Tmax}),
thus the fitting process is needed only for the diffusion barrier.
An additional parameter is the width of the narrow Gaussian distribution
of the energy barriers for each of the three types of 
adsorption sites.
The energy barriers for desorption of molecules from the different
types of adsorption sites and the population ratios of these sites
are presented in Tables 2 and 3, respectively.

\subsection{Analysis of Experiments with Irradiation of Atoms}
\label{subsec:Irradiation with Atoms}

In the experiments with irradiation of H and D atoms
(H+D for brevity) or of D atoms only (D+D), 
the adsorbed atoms diffuse on the surface and recombine.
The experimental TPD curves 
for the HD and D$_2$ molecules
can be fitted using either the
simple or the complete rate equation models.
The parameters that characterize the behavior
of hydrogen molecules on the surface, 
namely the desorption barriers
$E_{\rm H_2}^{\rm des}(j)$
and the population ratios
$\mu_j$,
are available from
the experiments with molecular irradiation.
Thus, the only fitting parameters that remain are
the energy barriers for diffusion and desorption of hydrogen atoms,
namely
$E_{\rm H}^{\rm diff}$ 
and 
$E_{\rm H}^{\rm des}$.
In Figs. 2 and 3 one observes that the two peaks
in the TPD curves obtained after irradiation with H+D on LDI 
coincide with the two higher peaks obtained after
irradiation with HD molecules.
Similarly, the peak obtained after irradiation of H+D on HDI
coincides with the peak obtained after irradiation of HD 
molecules (Fig. 5).
The location of this peak is determined by the desorption barriers
of HD molecules, no matter whether these molecules were deposited or
formed on the surface.

In Fig. 3 we present five TPD curves for H+D irradiation
on LDI, for different irradiation times.
In each curve one can identify two peaks.
The location of the high temperature peak is 
independent of the irradiation time.
This is an indication of first order kinetics, namely that
the molecules desorbed at this stage were formed at a lower
temperature. 
They remained on the surface until the temperature became sufficiently
high for them to desorb.
The low temperature peak in each curve in
Fig. 3 shifts slightly to the right
as the irradiation time is reduced.
This is a feature of second order kinetics,
indicating that within this peak the molecules
were desorbed upon formation. 
Our interpretation is that these molecules were formed
in shallow adsorption sites and thus easily desorbed at
the temperature in which they recombined.
These shallow adsorption sites can be identified with those
that correspond to the leftmost among the three peaks 
in the molecular irradiation experiment shown in Fig. 2.
The fact that the same set of parameters can fit all the
curves in Fig. 3, obtained for different irradiation times, 
as well as all the other experiments on LDI,
provides very strong evidence in favor of our models.
Similarly, a single set of parameter provides very good
fits to all the experiments on HDI.
These parameters are presented in Tables 2 and 3.

\subsection{Implications to Interstellar Chemistry}

Using the parameters obtained from the experiments we now
calculate the recombination efficiency of H$_2$ molecules on
ice surfaces under interstellar conditions.
The recombination efficiency is defined as the fraction
of hydrogen atoms adsorbed on the surface which come out
as molecules.
In Fig. 8 we present the recombination efficiency of H$_2$
molecules vs. surface temperature for LDI under flux of
0.73$\cdot 10^{-8}$ ML s$^{-1}$.
This flux is within the typical range for both 
diffuse and dense interstellar clouds. 
This particular value corresponds to a density of H atoms in
the gas phase of 10 (atoms cm$^{-2}$), gas temperature
of 100 K and $5 \times 10^{13}$ adsorption sites per
cm$^{2}$ on the surface
\citep{Biham2001}.

A window of high efficiency is found between 11-16K.
At higher temperatures atoms desorb from the surface before
they have sufficient time to encounter each other.
At lower temperatures diffusion is suppressed while
the Langmuir rejection leads to saturation of the surface
with immobile H atoms and recombination is suppressed.
At these low temperatures, the steady-state coverage is high
and the rate equation model may not apply.
As already suggested in Pirronello et al. (1999),
mechanisms such as Eley-Rideal or diffusion by tunneling, which are
not taken into account in the model,
may become significant in this regime.
Nevertheless, we can speculate that if the Langmuir rejection remains
significant even at
higher coverage, the trend in recombination efficiency
shown should remain qualitatively correct.
Luckily, such low temperatures are rarely of astrophysical
interest. At high temperature atoms desorb from
the surface before they have sufficient time to recombine, 
thus inhibiting hydrogen
recombination at high temperature.
For both high and low density ice the
calculated asymptotic values of coverage 
are very low and well within the regimes of
experimentation and subsequent numerical
simulation. 
Consequently, the relevance of the model is justified.
Our results indicate that recombination efficiency of hydrogen on both
HDI and LDI is high in the temperature range of $\simeq$14-20 K which is
relevant to the interstellar clouds. Therefore
these ice surfaces seem to be good candidates 
for interstellar grain components
on which hydrogen may recombine with high efficiency.

\section{Discussion}
\label{sec:discussion}

One of the important features demonstrated by these experiments is the
relevance of the surface morphology to the recombination rate of H$_2$
molecules. 
Surfaces of amorphous ice are difficult to model due to the paucity
of morphological information on either actual ISM ice coated grains
or their laboratory analogues.
This difficulty gives rise to some uncertainty
about the role of quantum effects. 
In principle, quantum tunneling of H atoms on the surface
should be considered. 
However, these experiments, much like
earlier experiments on amorphous carbon and polycrystalline olivine
samples
\citep{Pirronello1997b}, 
indicate 
that the mobility of the hydrogen atoms is very low at low temperatures.
Thus, the effect of tunneling appears to be small.
Cazaux and Tielens showed that quantum effects might be important 
at high coverage.
However, at the low coverages of H atoms on
interstellar grains, tunneling effects 
are expected to be very small as can be seen
in  Fig. 6b in Ref. 
\citep{Cazaux2004},
obtained for
low irradiation (exposure) times.

The effect of the ice morphology is best seen in the difference between
the types of adsorption sites and their depth. 
The more porous structure
of the LDI 
is reflected in the broad
distribution of the 
activation energies for desorption from different adsorption sites
which spans the range  between 40 to 60 meV
\citep{Jenniskens1995}.
In comparison, the energy barriers for desorption from HDI exhibit a
narrow distribution around 68.7 meV.
The energy barriers for diffusion 
of H atoms on the ice surfaces are found to be higher than
those obtained earlier for more smooth surfaces such as the
polycrystalline olivine analyzed by Katz et al.~(1999).
Thus, the 
mobility of H atoms on the highly amorphous ice is 
relatively low,
in qualitative agreement with the findings of Smoluchowski 
(1983).
The energy barriers for diffusion and
desorption of H atoms and molecules on amorphous ice
were calculated 
using a detailed model of the amorphous ice structure
and the interactions at the atomic scale
\citep{Buch1991a,Buch1991b,Hixson1992}.
The energy barriers for H desorption
were found to be distributed
between 26 to 57 meV,
while the 
barriers for diffusion were found to be between 13 to 47 meV. 
Our results for LDI are in good agreement with 
these calculations, while 
for HDI our barriers are somewhat higher 
(see Tables 2 and 3).
For $H_2$ molecules they find a distribution of 
activation energy barriers for desorption in the range between
400 and 1200 K, namely 33 meV - 133 meV. Our three barriers
for the desorption of HD molecules are in the range of 46 - 61 meV,
namely around the center of the calculated distribution.
Thus, our model captures the effects of the amorphous ice
structure and interactions, in agreement with detailed calculations
at the atomic scale.

In the experiments we use a combination of the H and D isotopes.
In most cases the production rate of HD is measured, 
complemented by some measurements of D$_2$ formation.
Experiments measuring H$_2$ formation
are difficult to carry out.
This is due to the 
existence of background H$_2$ which is the most abundant residual gas
in the well baked ultra-high vacuum chamber.
Even the small percentage of undissociated molecules that are sent
with the atomic beam can influence the results.
The use of hydrogen isotopes enables us to 
perform measurements with much more accuracy,
and obtain the important physical parameters.
The extrapolation of our results to processes 
that involve only H and H$_2$ is non-trivial. 
However, the experiments in which the formation of 
HD and D$_2$ were analyzed, 
enable us to examine
the isotopic differences. 
The analysis shows that variations
between the energy barriers involved in the HD and D$_2$ 
production processes do exist. 
Nonetheless, the two processes are qualitatively similar and
the only difference is in a
small increment in the energy barriers of
the D$_2$-formation process in comparison to the HD.
We thus conclude that the
energy barriers we obtain
provide a good approximation to those involved in the
production of H$_2$ on ice surfaces.

Previous rate equation models, used in analysis of hydrogen
recombination on carbon and olivine 
\citep{Katz1999,Cazaux2002,Cazaux2004},
assumed that a significant fraction 
(denoted by $1-\mu$)
of the hydrogen molecules,
desorb from the surface upon formation
(note that $\mu$ in these three papers had a different meaning
than in the present paper).
The rest of the molecules (namely, a fraction $\mu$ of them)
remain on the surface and desorb thermally at a later stage.
The origin of the parameter $\mu$ remained unclear, but it
was necessary in order to fit the data.

The model used in the present paper provides a physical motivation
for the parameter $\mu$, based on a different interpretation
of the fact that part of the molecules quickly desorb.
It is based on the broad
distribution of the energy barriers for desorption of
hydrogen molecules which exhibits three types of 
adsorption sites.  
According to this model 
the recombined molecules equilibrate with the surface and
reside in adsorption sites until thermal desorption takes place.
Molecules formed in the most shallow adsorption sites 
desorb quickly because
by the time they are formed, the surface temperature
is already high enough to activate their thermal desorption.
The fraction, $\mu_1$ of the shallow adsorption sites among all
the molecular adsorption sites is obtained from the experiments
in which molecules are irradiated.

As mentioned above, in the models used here it is assumed
that hydrogen molecules do not desorb immediately upon formation.
Instead, they stay trapped in the adsorption sites or diffuse
between them until thermal desorption takes place.
Consequently, one needs to consider mechanisms for the dissipation
of the excess energy acquired from the recombination process
in order to prevent prompt desorption. 
Although classical molecular
dynamics simulations for H$_2$ formation on amorphous ice performed by
Takahashi et al. (1999) give a large average
kinetic energy ($\simeq$530 meV) to the desorbing molecules,
both Roser et. al. (2003) and independently 
Hornekaer et al. (2003) showed experimentally
that this kinetic energy is much smaller ($\simeq$3 meV), in agreement
with our assumption that the excess energy is dissipated elsewhere.
There are various mechanisms for efficient heat transfer from the molecule
to the surface. 
One possible mechanism
may be due to the very irregular structure of the sample surfaces. 
Both LDI and HDI surfaces are disordered.
Even if molecules are promptly released upon formation,
they do not necessarily go directly into the vacuum
but they might undergo a multiple series of collisions in which part of their
energy is released to the solid. 
These multiple hits might lead to a subsequent 
re-adsorption of the molecule on the ice surface.
Such a mechanism was already identified for H and D atoms impinging on an
amorphous ice particle 
\citep{Buch1991b}.

The experiments done by our group as well as by Hornekaer et al.
have explored several types of ice morphology.
The experiments analyzed here were done on amorphous ice
of more than 1000 ML, where we have used two different ices, LDI and HDI
(both amorphous and porous). 
The experiments by Hornekaer et. al. 
\cite{Hornekaer2003}
were done on ice that was prepared 
using the same procedure used by our group to produce the HDI, 
with 2000 ML thickness.
Their atom beam fluxes were
$\sim$ $10^{13}$
(atoms cm$^{-2}$ sec$^{-1}$), namely
about one order of magnitude higher than the
fluxes used in our experiments.
The range of exposure times was comparable.
Hornekaer et al.
investigated the kinetics of HD formation and measured
the efficiency of recombination and the energetics of
the molecules released from the ice layer after formation.
The efficiency values they obtained are close to those obtained by our group
on amorphous ice.
The energy distribution of molecules formed showed
that at least in porous amorphous ice, molecules are
thermalized by collisions with the walls of pores (where they are
formed)
before they emerge into the gas phase.
Hornekaer et al. performed
TPD experiments in which they irradiated H and D
atoms either simultaneously or sequentially
after waiting a delay time interval before
dosing the other isotope.
On porous ice the results they obtained are
consistent with a recombination occurring quickly after atom dosing
due to a high mobility of the adsorbed atoms
even at temperature as low as 8 K.
This high mobility was attributed either to quantum
mechanical diffusion or to the so called hot atom mechanism,
where thermal activation is not likely to play a significant
role at this temperature.
This conclusion is sensible in light of the
high coverages of H and D atoms irradiated in their experiments,
which required the adsorbed atoms to diffuse only short distances before
encountering each other.
As already suggested by 
Pirronello et al. (2004a,b) and Vidali et al. (2005),
the hot atom mechanism may be able to provide the required mobility.
In this case H and D atoms retain a good
fraction of their gas phase kinetic
energy during the accommodation process
\citep{Buch1991b,Takahashi2001}.
This enables them
to travel on the ice surface and inside its pores
for several tens of Angstroms exploring several adsorption sites and
recombining upon encountering already adsorbed atoms.
However, in our experiments, because of the low
coverage, the number of sites explored by the hot atom is not large
enough to encounter an adsorbed atom and react with it with significant
probability 
\citep{Vidali2005}.
Thus, there is no contradiction between the interpretations that
molecular hydrogen formation is dominated by
the hot atom mechanism at high coverage and by
thermally activated mobility of the adsorbed H atoms at low coverage.

In addition, Hornekaer et al. studied non-porous ice with only 20 ML. 
In this paper 
we have not analyzed such non-porous samples and therefore cannot
provide any comparison in this case. 
On the other hand, we have used LDI for which 
Hornekaer et al. (2003)
do not provide analogous results that enable a comparison. 
Our results and analysis show that on LDI 
the recombination of H atoms is thermally activated, 
and it is not efficient at low temperatures.

In their analysis, Cazaux and Tielens (2004) 
considered a rate equation model 
with a single energy barrier for molecular desorption from carbon and
olivine surfaces. 
In their model
chemisorption
sites play a role in the
recombination process of hydrogen molecules and each
hydrogen isotope is treated differently.
Although chemisorption sites might play a role in the recombination process,
these sites do not seem to play an important role, at least not at the
conditions explored in our experiments 
(low temperatures and low coverage as expected for 
interstellar dense cloud environments). 
However, the analysis of Cazaux and Tielens
may provide a clue on H$_2$ formation in
photon dominated regions. 
We have not treated each atomic isotope differently but instead we used
different effective averaged values for the experiments involving H and D
atoms and those involving only D atoms. 
Consequently obtained
a model that requires fewer fitting parameters.

\section{Summary}
\label{sec:Summary}

Experimental results on the formation of molecular
hydrogen on amorphous ice under
conditions relevant to interstellar clouds were analyzed using rate
equation models.
By analyzing the results of TPD experiments and fitting them to rate
equation models, the essential parameters of the 
process of molecular hydrogen formation on ice surfaces were obtained.
These parameters include the activation energy barriers for
diffusion and desorption of hydrogen atoms and molecules on the
ice surface. 
While we identify only one type of adsorption site for hydrogen 
atoms, three types of adsorption sites are found for molecules
on LDI,
with different activation energies.
The parameters that determine what fraction of the molecular
adsorption sites belong to each type are also found.
Our model enables a unified description of
several first and second order processes that involve 
irradiation by either hydrogen atoms or molecules,
all within the framework of a single model.

The rate equation model allows us to extrapolate the production
rate of hydrogen molecules from
laboratory conditions to astrophysical conditions.
It thus provides a quantitative evaluation of the
efficiency of various ice surfaces as catalysts in 
the production of hydrogen molecules in interstellar clouds.
It is found that the production efficiency strongly depends on
the surface temperature.
Both types of ice samples studied here exhibit 
high efficiency within a range of surface temperatures
which is relevant to dense molecular clouds which include 
ice-coated dust grains. 

\section{Acknowledgments}
\label{sec:Acknowledgments}

This work was supported by the Adler Foundation for Space Research
and the Israel Science Foundation (O.B),
by NASA through grants 
NAG5-11438 and NAG5-9093 (G.V),
and by the Italian Ministry for University and Scientific Research
through grant 21043088 (V.P).

\clearpage
\newpage

\begin{table}
\begin{center}
\caption{
List of the
TPD experiments on low and high density amorphous ice
analyzed in this paper.}
\begin{tabular}{|c|c|c|c|}
\tableline\tableline\hline\hline
\rm Ice & Atoms/Molecules & Irradiation Time  & Irradiation \\
   Type & Type            &  (minutes)        & Temperature (K) \\ 
\hline \hline
     LDI & H + D atoms & 2  & $\simeq$9.5 \\
     LDI & H + D atoms & 4  & $\simeq$9.5 \\
     LDI & H + D atoms & 8  & $\simeq$9.5 \\
     LDI & H + D atoms & 12 & $\simeq$9.5 \\
     LDI & H + D atoms & 18 & $\simeq$9.5 \\
     HDI & H + D atoms & 4  & $\simeq$14.5  \\
     HDI & H + D atoms & 6  & $\simeq$14.5  \\
     HDI & H + D atoms & 8  & $\simeq$14.5  \\
     HDI & H + D atoms & 18 & $\simeq$14.5 \\
     LDI & D atoms & 4      & $\simeq$10 \\
     LDI & HD molecules & 4 & $\simeq$9 \\
     HDI & HD molecules & 4 & $\simeq$15 \\
     LDI & D$_2$ molecules & 4 & $\simeq$10 \\
     HDI & D$_2$ molecules & 4 & $\simeq$15 \\ \hline
\end{tabular}
\end{center}
\end{table}

\clearpage
\newpage

\begin{table}
\begin{center}
\caption{
The energy barriers obtained by the
fitting of the TPD curves for
low and high density ice surfaces.
$E_{\rm H}^{\rm diff}$ 
is the barrier for
atomic diffusion,
$E_{\rm H}^{\rm des}$ 
is the barrier
for atomic desorption and 
$E_{\rm H_2}^{\rm des}(j)$, $j=1,2,3$ are
the barriers for molecular desorption from
sites of type $j$.}
\begin{tabular}{|c|c|c|c|c|c|c|}
\tableline\tableline\hline\hline
\rm Material & Molecule & {\em $E_{H}^{\rm diff}$ ({\rm meV})}
& {\em $E_{\rm H}^{\rm des}$ ({\rm meV})}
& {\em $E_{\rm H_2}^{\rm des}(1)$({\rm meV})}
& {\em $E_{\rm H_2}^{\rm des}(2)$({\rm meV})}
& {\em $E_{\rm H_2}^{\rm des}(3)$({\rm meV})} \\
& Type & & & & &       \\ \hline \hline
Low Density & HD & 44.5  & 52.3 & 46.5 & 52.8 & 61.2 \\
Ice&$D_2$ & 41.0  & 45.5 & 40.7 & 53.3 & 65.5 \\ \hline
High Density& HD & 55 &  62 & 68.7 \\
Ice&$D_2$ & & & 72.0 \\ \cline{1-5}
\end{tabular}
\end{center}
\end{table}

\begin{table}
\begin{center}
\caption{
The parameters of the population ratio obtained by the
fitting of the TPD curves for
low and high density ice surfaces.
$\mu_{1}$, $\mu_{2}$ and $\mu_3$ are the population ratios
of the different adsorption sites of the molecules and $\sigma$
is the standard deviation parameter for the energy barriers's Gaussian
distribution around the characteristics energy barriers for molecules
desorption.}
\begin{tabular}{|c|c|c|c|c|c|}
\tableline\tableline\hline\hline
\rm Material & Molecule
& {$\mu_{1}$}
& {$\mu_{2}$}
& {$\mu_{3}$}
& {$\sigma$ (meV) } \\
 &Type&&&&\\ \hline \hline
Low Density & HD & 0.2 & 0.47 $ \pm $ 0.075& 0.33 $ \pm $ 0.075 &0.0045 \\
Ice&$D_2$ & 0.23  $ \pm $ 0.05 & 0.51 $ \pm $ 0.05  & 0.26  & 0.0057\\ \hline
High Density &HD & 1& & &0.006 \\
Ice&$D_2$ &1 & &&0.006 \\  \hline
\end{tabular}
\end{center}
\end{table}

\clearpage
\newpage

\begin{figure}
\includegraphics[width=5in]{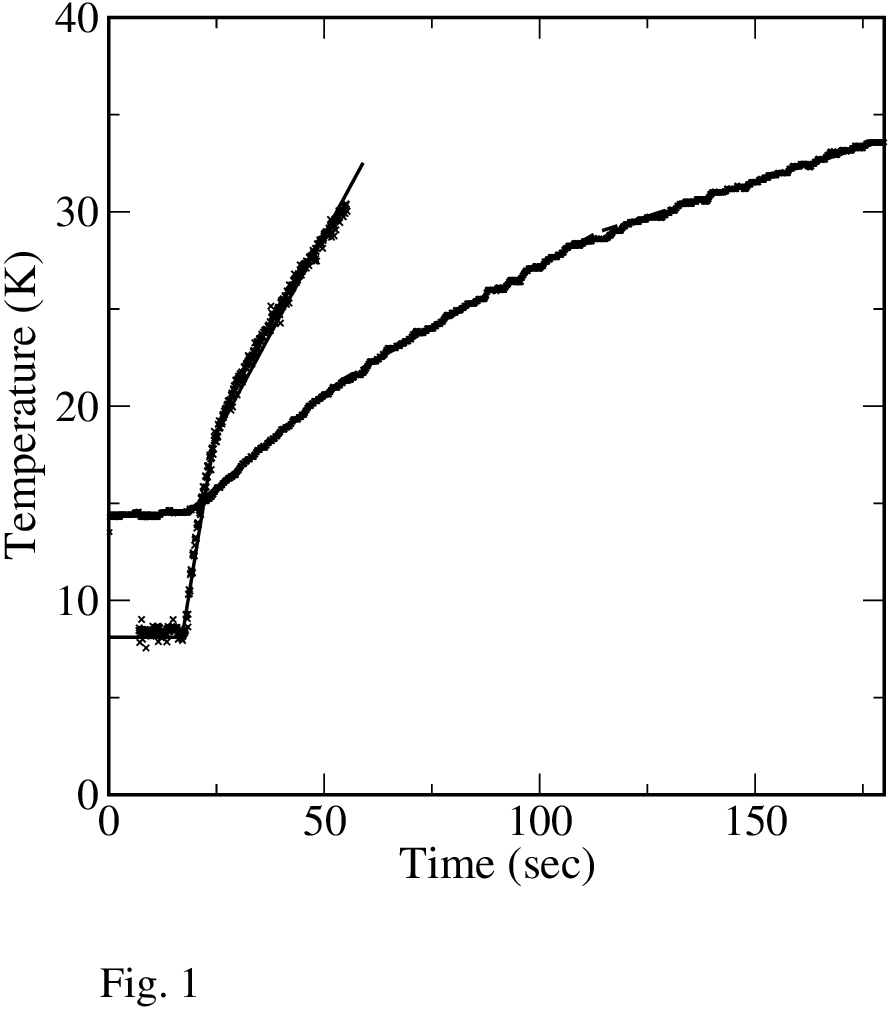}
\caption{
The time dependence of the surface temperature during typical
temperature ramps 
in TPD experiments on
LDI (steeper line)
and
HDI (less steep line).
The symbols show the experimental measurements
and the lines are piecewise linear fits.
The irradiation phase is not shown. 
}
\end{figure}

\begin{figure}
\includegraphics[width=5in]{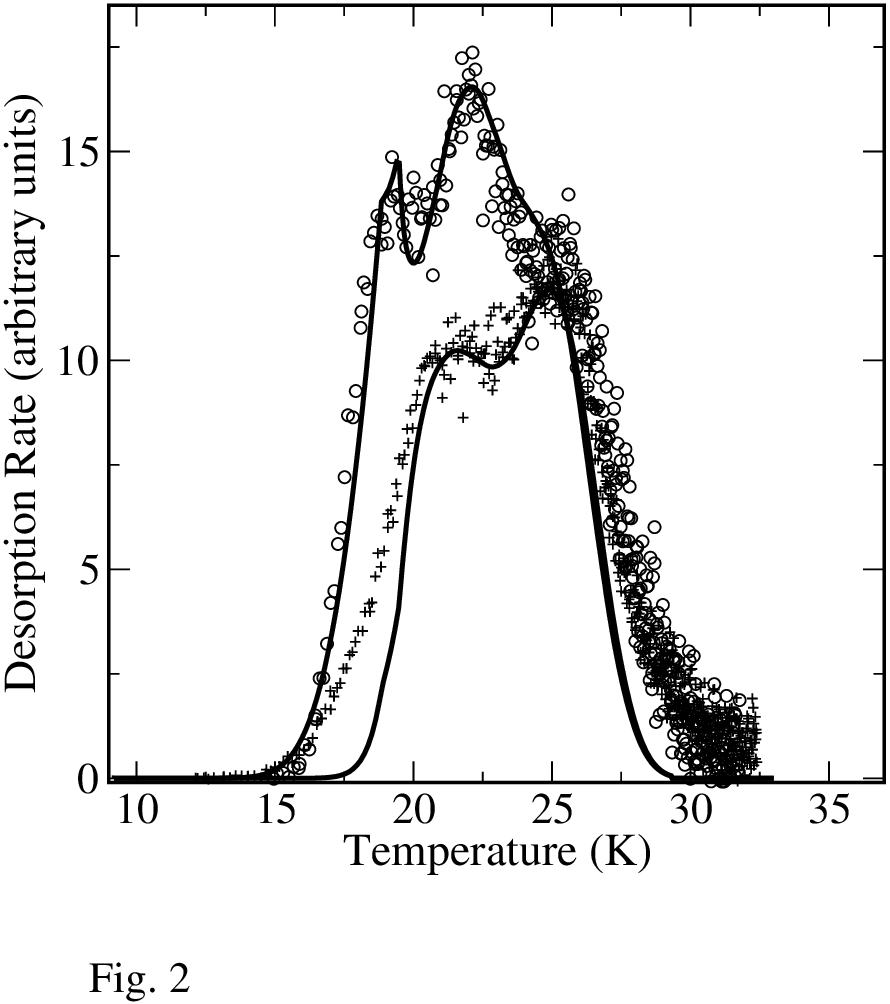}
\caption{
TPD curves of HD desorption after irradiation with
HD molecules ($\circ$) and H+D atoms (+) 
on low density ice.
The irradiation time is 4 minutes.
The solid lines are fits obtained by the 
complete rate equations model.
}
\end{figure}

\begin{center}
\begin{figure}
\includegraphics[width=5in]{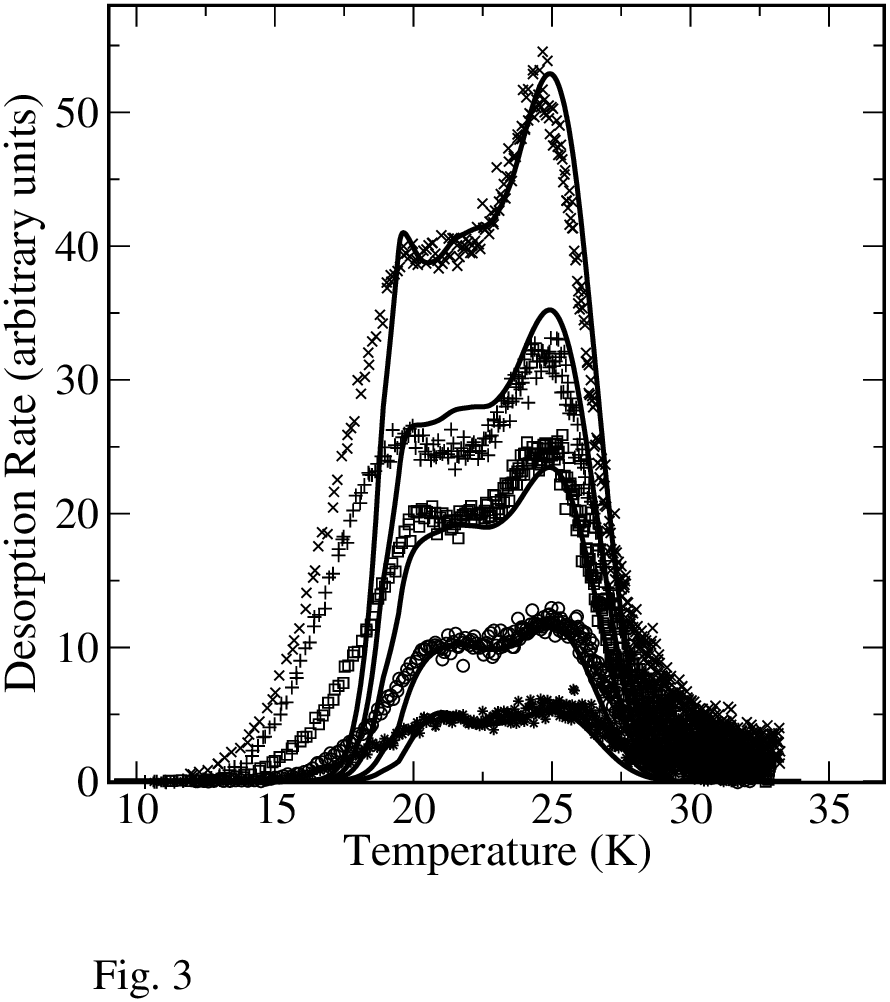}
\caption{
TPD curves of HD desorption after irradiation with
H+D atoms on LDI.
The irradiation times are 2 ($\ast$), 4
($\circ$), 6 ($\Box$) 12 (+) and 18 ($\times$) minutes. 
The solid lines are fits obtained by the 
complete rate equation model.
}
\end{figure}
\end{center}

\begin{center}
\begin{figure}
\includegraphics[width=5in]{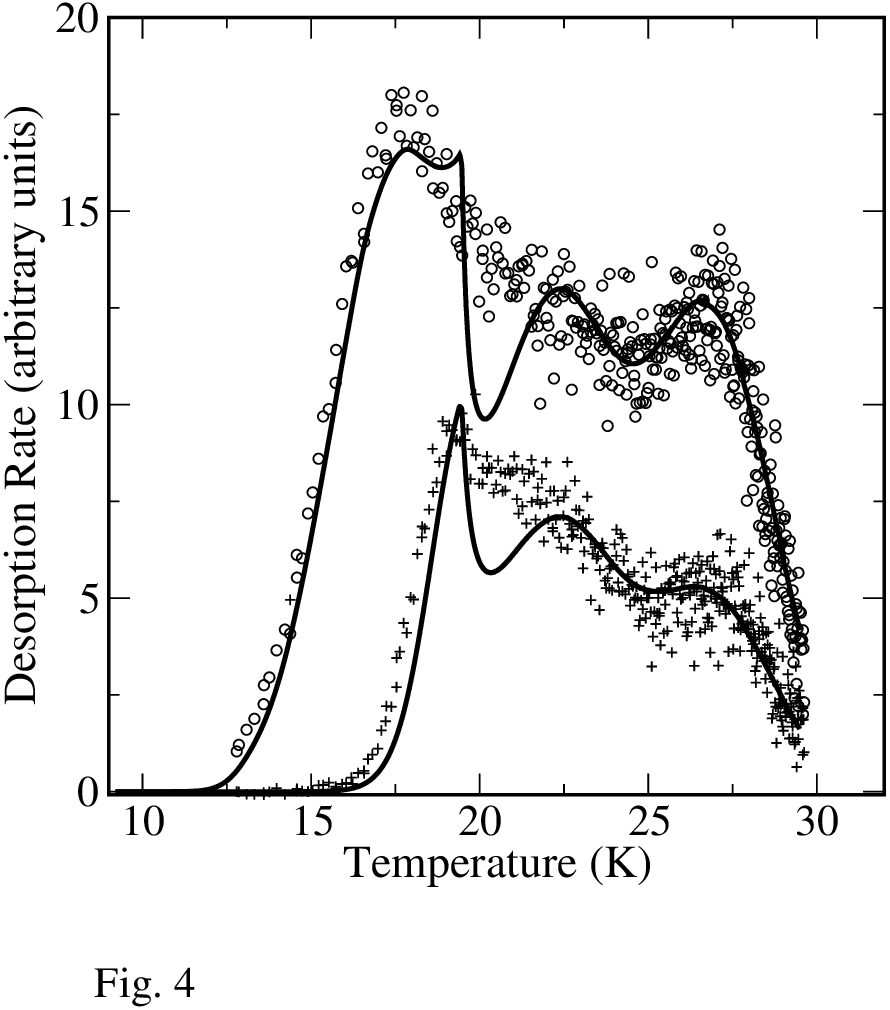}
\caption{
TPD curves of D$_2$ desorption after irradiation with 
D$_2$ molecules ($\circ$) and D+D atoms (+)
on LDI,
fitted by the
complete rate equation model
(solid lines). 
The irradiation time is 4 minutes. 
}
\end{figure}
\end{center}

\begin{center}
\begin{figure}
\includegraphics[width=5in]{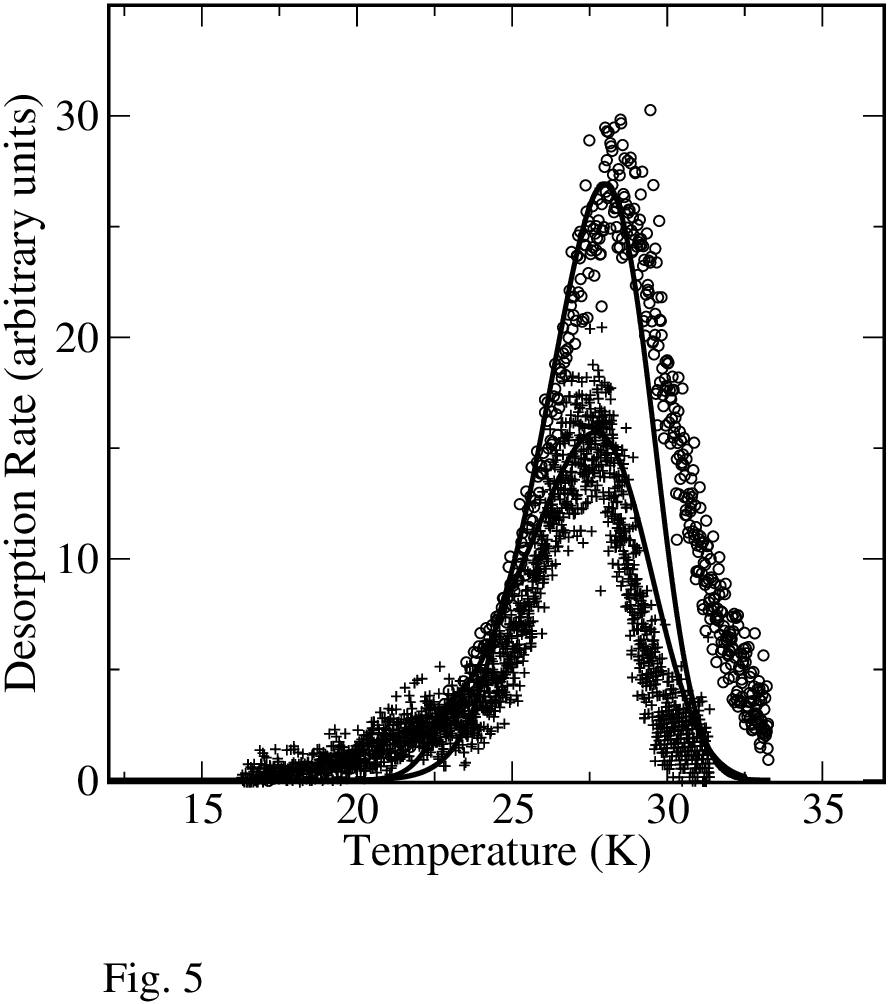}
\caption{
TPD curves of HD desorption after irradiation with
HD molecules ($\circ$) and H+D atoms (+) 
on high density ice,
fitted by the 
complete rate equations model
(solid lines).
Irradiation time is 4 minutes. 
}
\end{figure}
\end{center}

\begin{center}
\begin{figure}
\includegraphics[width=5in]{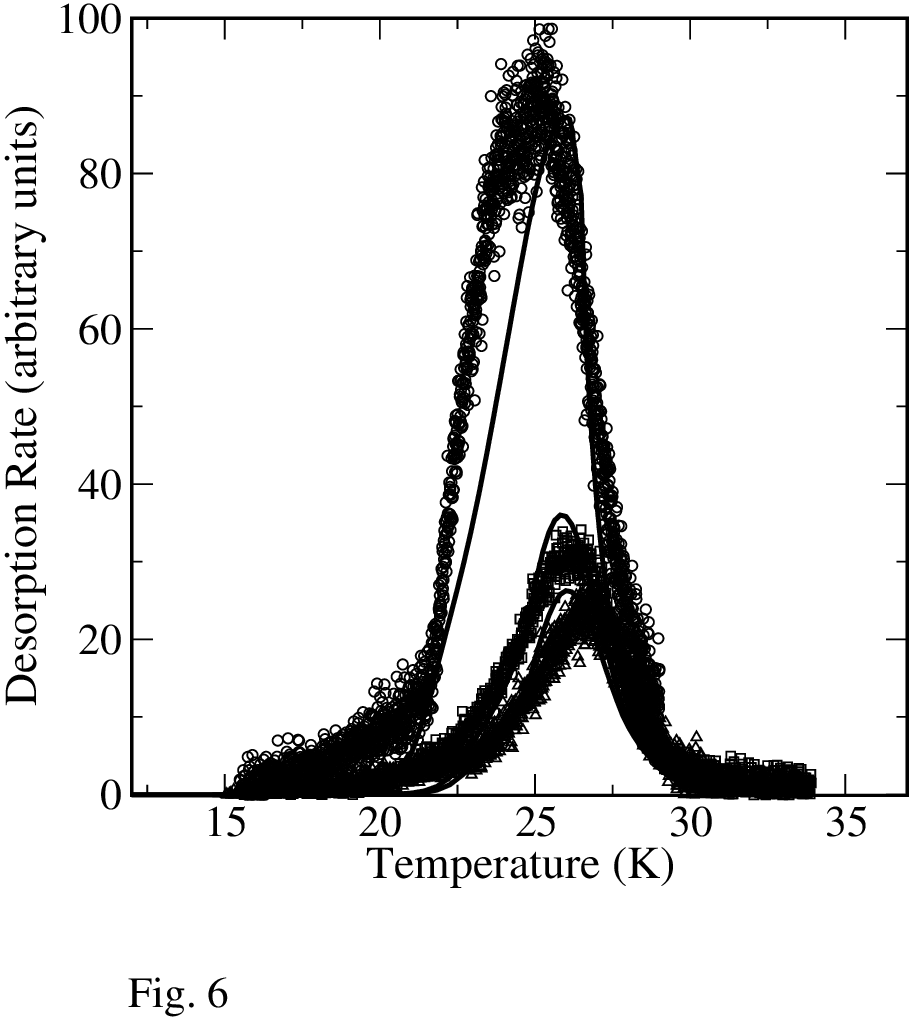}
\caption{
TPD curves of HD desorption after irradiation with
H+D atoms on HDI, 
fitted by the 
complete rate equations model
(solid lines).
The irradiation times are 6 ($\triangle$),
8 ($\Box$) and 18 ($\circ$). 
}
\end{figure}
\end{center}

\begin{center}
\begin{figure}
\includegraphics[width=5in]{f7}
\caption{
TPD curves of D$_2$ desorption after irradiation with
D$_2$ molecules on HDI ($\circ$),
fitted by the complete rate equations model
(solid line).
The irradiation time is 4 minutes. 
}
\end{figure}
\end{center}

\begin{center}
\begin{figure}
\includegraphics[width=5in]{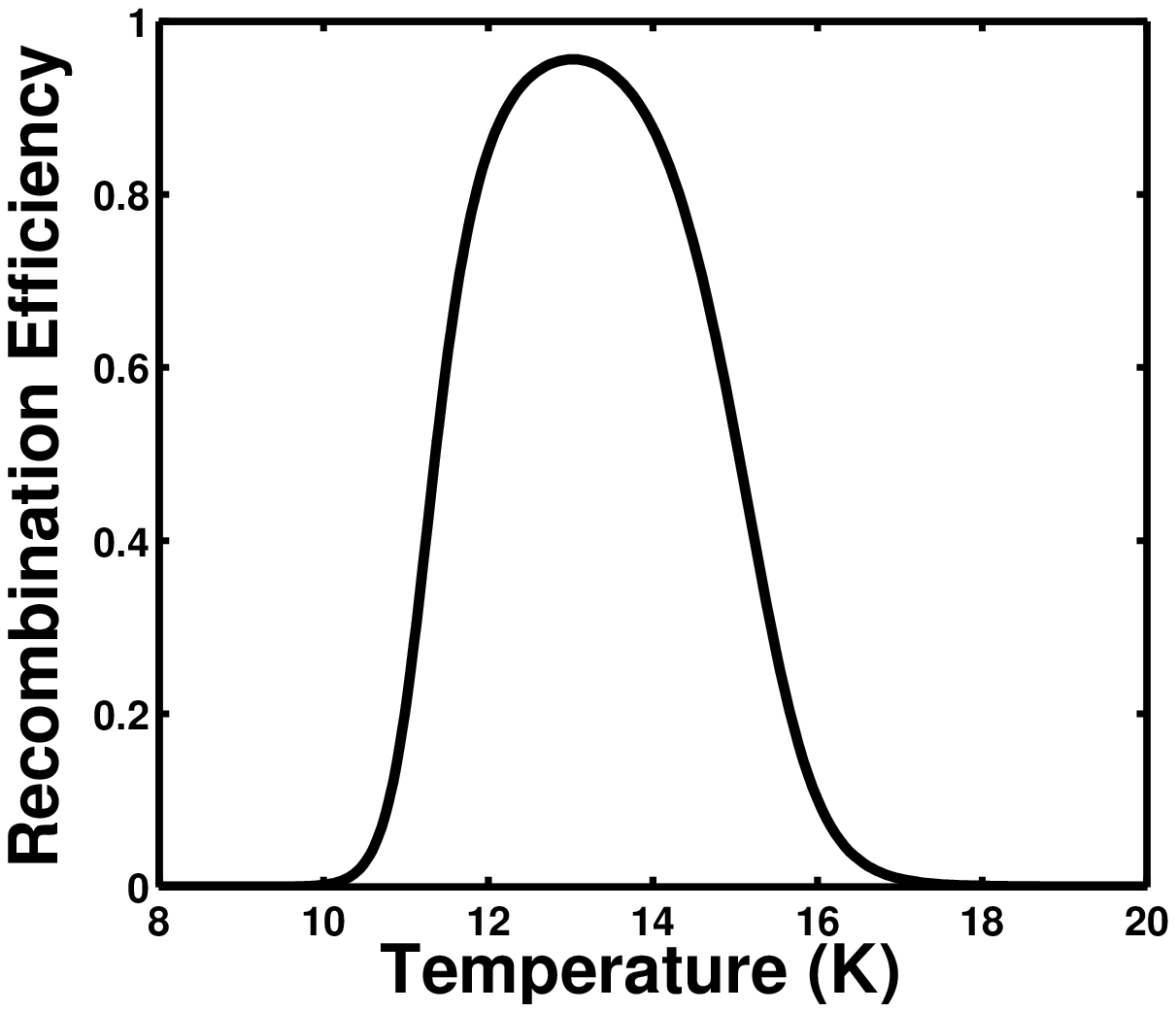}
\figcaption{
Recombination efficiency of molecular hydrogen at steady state on LDI
as a function of the temperature T (K), using the parameters
obtained from experimental measurements of HD desorption curves.
The fluxes of hydrogen atoms is 0.73$\cdot 10^{-8}$ ML s$^{-1}$.
}
\end{figure}
\end{center}

\begin{center}
\begin{figure}
\includegraphics[width=5in]{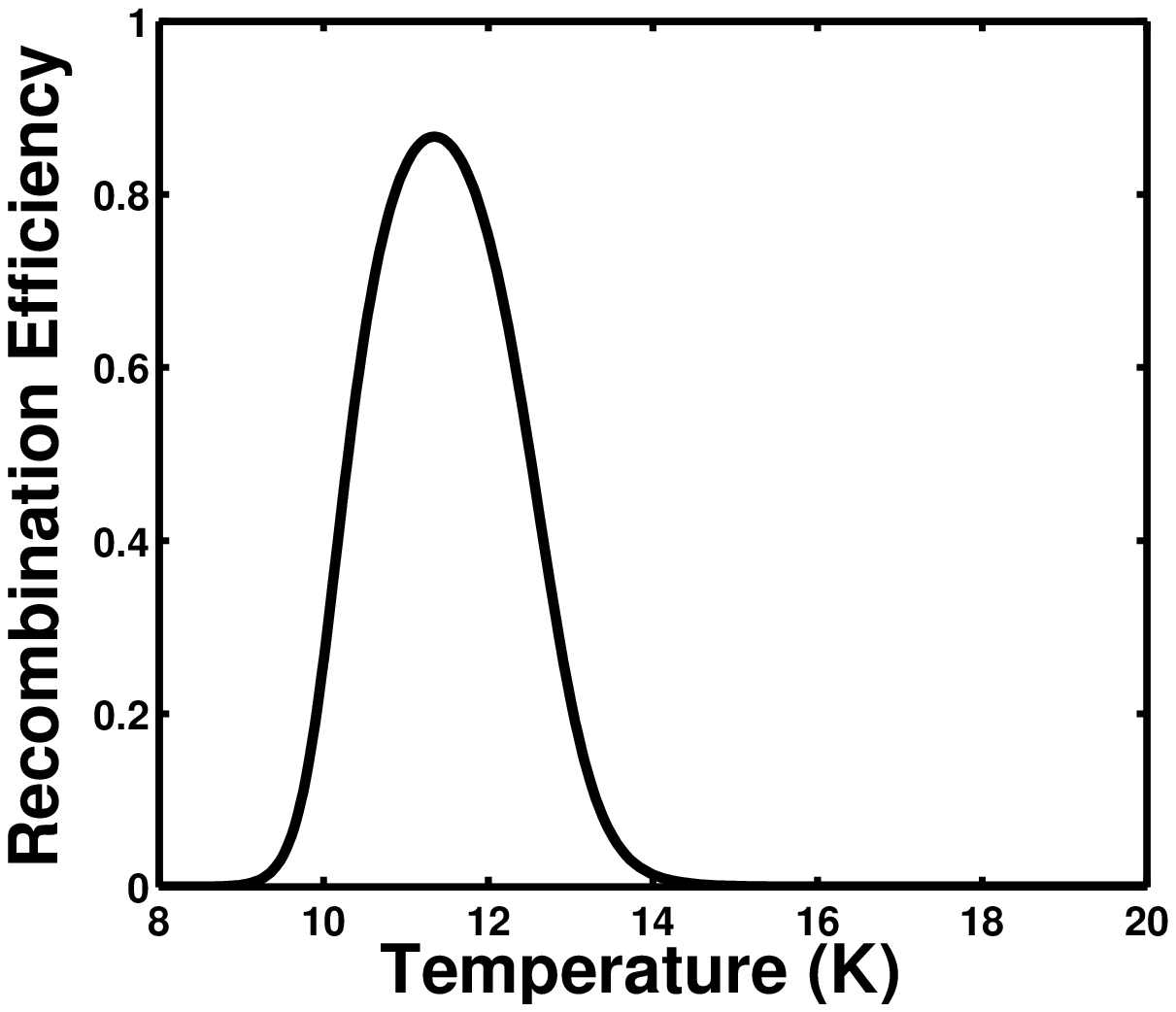}
\figcaption{
Recombination efficiency of molecular hydrogen at steady state on LDI
vs. the temperature T (K), using the parameters obtained from experimental
measurements of D$_2$ desorption curves.
The flux is the same as in Fig. 8.
}
\end{figure}
\end{center}

\begin{center}
\begin{figure}
\includegraphics[width=5in]{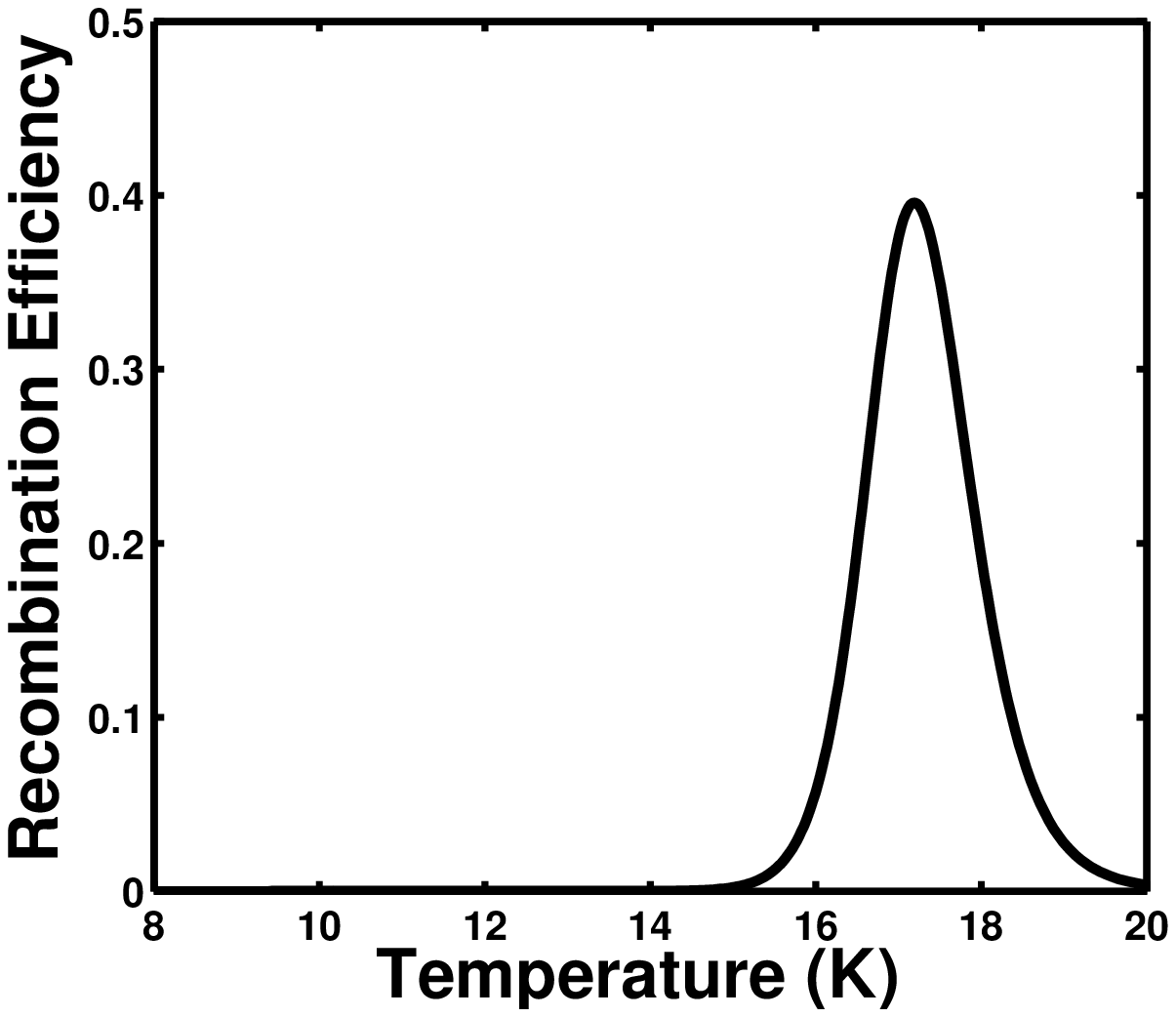}
\figcaption{
Recombination efficiency of molecular hydrogen at steady state on HDI
vs. the temperature T (K), using the parameters obtained from 
experimental measurements of HD desorption curves.
The flux is the same as in Fig. 8.
}
\end{figure}
\end{center}
\end{document}